\documentclass[aip,reprint]{revtex4-1}
\usepackage{graphicx}
\usepackage{dcolumn}
\usepackage{bm}
\usepackage{color}
\newcommand{\la}{\raise0.3ex\hbox{$<$}\kern-0.75em{\lower0.65ex\hbox{$\sim$}}}
\newcommand{\ga}{\raise0.3ex\hbox{$>$}\kern-0.75em{\lower0.65ex\hbox{$\sim$}}}

\begin{document}


\title{Using Collisions of AGN Outflows with ICM Shocks as Dynamical Probes} 



\author{T. W. Jones}
\email{twj@umn.edu}
\affiliation{School of Physics and Astronomy, University of Minnesota, Minneapolis, MN}
\author{Chris Nolting}
\affiliation{School of Physics and Astronomy, University of Minnesota, Minneapolis, MN}
\author{B. J. O'Neill}
\affiliation{School of Physics and Astronomy, University of Minnesota, Minneapolis, MN}
\author{P. J. Mendygral}
\affiliation{Cray, Inc.}
\affiliation{School of Physics and Astronomy, University of Minnesota, Minneapolis, MN}


\date{\today}

\begin{abstract}
In this paper we lay out a simple set of relationships connecting the dynamics of
fast plasma jets to the dynamical state of their ambient media. The objective is
to provide a tool kit that can be used to connect the morphologies of radio
AGNs in galaxy clusters to the dynamical state of the local ICM. 
The formalism is intended to apply to jets whether they are relativistic or 
non-relativistic. Special attention is paid to interactions involving ICM shocks, 
although  the results can be applied more broadly. Our formalism emphasizes the
importance of the relative Mach number of the impacting ICM flow and
the internal Mach number of the AGN jet in determing how the AGN outflows evolve.

\end{abstract}


\maketitle 

\section{Introduction}
\label{intro}
Clusters of galaxies are the last and most massive class of bound structures to
form out of the Big Bang ($z \la 1, R \sim \rm{Mpc},~M\sim 10^{14}-10^{15}\rm{M}_{\odot}$). Cluster assembly combines irregular accretion
from surrounding diffuse matter, especially from filaments, and violent mergers with other clusters. As in the universe
at large, most cluster  matter is non-baryonic, ``dark matter''. The baryonic
matter within clusters is predominantly hot, diffuse plasma ($kT ~\ga~ keV, n_e~ \la~ 10^{-2}\rm{cm}^{-3}$); the intracluster medium (ICM). The ICMs are shocked and
stirred throughout their formation \cite{krav12,vazz16,zing16}. The resulting ICM flow structures provide vital 
information about the large scale dynamics of cluster formation, as well as about the
physics of the ICM plasma\cite{schek06}. Cluster-scale shocks, in particular, are tell-tale
consequences of
merger events \cite{mark07}. Tracing and deciphering these ICM flow structures are essential
to our understanding of cluster formation. 

While thermal X-rays have successfully revealed some 
ICM shocks and other flow features \cite{mark07,walker,hitomi}, distinct X-ray signatures are often subtle and hard to isolate and measure cleanly.
Observations of the Sunyaev-Zeldovich effect
add some statistical information about the ICM pressure and velocity distributions\cite{adam15}.
But, a full, clear picture requires additional, complementary tools. 
Thankfully, non-thermal synchrotron emissions from $>$ GeV electrons (CRe) in the ICM and
in embedded objects  may help reach this goal.

So far, the principal non-X-ray tool in efforts to characterize ICM dynamics has been cluster-scale, diffuse radio emissions (radio ``halos'', ``mini-halos'' and ``relics'') by electrons  within the ICM
that are energized or re-energized by cluster shocks and/or turbulence \cite{bj14}. 
There are multiple plausible origins for those diffuse CRe,
which in various models may be recent or ancient \cite{pinzk15,kr15}. Their links to local, current  ICM dynamics result from the
fact that they have rather short lifetimes against synchrotron losses and
inverse Compton scattering of CMB photons ($<$ 100 Myr).  Observed CRe must be ``energetically younger'' than this. Since ICM flow velocities are
characteristically $\sim 1$ kpc/Myr, such embedded CRe must have been
energized relatively close by on cluster scales. Although these emissions 
place essential constraints on ICM physics, their translation 
into distinct and firm ICM dynamical properties has so far proven to be quite
challenging \cite{kr15,don16}.

At the same time, radio-bright, bipolar plasma outflows from active galactic nuclei (radio AGN, henceforth, simply AGN) are
common in clusters \cite{ledlow96,mala16}. Interactions between those outflows and
ICM dynamical structures on scales of 10s to 100s of kpc offer
an additional and potentially very useful diagnostic probe of the ICM dynamics and
physics. That is quite separate from the increasing evidence that AGN-ICM feedback loops on those scales in the cores of
relatively undisturbed clusters are central to the properties of those ICMs \cite{li15}.
The focus of our discussion here, instead, is the consequences of AGN outflow encounters with 
cluster scale ICM flow
structures (especially shocks), and, in particular, impact on the consequent dynamical form of the AGN structures that develop as the AGN plasma penetrates
and generate cavities in the ICM. 

Specifically, we explore major distortions to the AGN flows
and the potential for using such distortions to identify
and characterize ICM flow structures, including shocks.
In subsequent publications we will
present analyses of high resolution 3D MHD simulations of AGN outflows in related dynamical ICM situations.
There we will examine details such as coherence and disruption of the AGN flows,
development of turbulence,
magnetic field evolution, as well as relativistic electron propagation, acceleration and emissions.
Our purpose now is more
restricted; namely it is to outline a simple, but coherent formalism that 
we have found to be useful in predicting behaviors of the above simulation experiments
and that
hopefully captures essential physics needed to interpret observed AGN 
interactions with dynamical ICMs. We note, in the context of the HEDLA workshop that
inspired this work, that many of the dynamical components laid out below should be
testable in laboratory experiments.

The best known and most widely discussed AGN distortion produced by asymmetric ICM interactions
are those
that have formed tails as a result of relative motions between the AGN host galaxy and
the surrounding ICM. So-called ``Narrow Angle Tailed'' (NAT) AGNs, in which the
originally collimated AGN outflows appear deflected into a pair of quasi-parallel
trails in the wake of the galaxy are especially common in disturbed clusters, where
relatively high velocity ICM motions are likely  \cite{rw68}. If there are apparent, but less dramatic deflections,
the AGNs are typically called ``Wide Angle Tailed'' (WAT) AGNs \cite{or76}.
Both are obvious candidates
as probes of ICM dynamics \cite{brb79,bliton98,sak00,mguda15}. The idea that ICM shocks encountering plasma bubbles or cavities
left behind by expired AGN might rejuvenate their CRe populations has also been widely
discussed \cite{eb02,pj11}. To the best of our knowledge, however, expected behaviors of
active AGN outflows colliding with ICM shocks has not been previously outlined,
although there have been several recent suggestions that observed NAT AGN may, in fact,
be associated with cluster radio relics \cite{bona14}. Here we do not address that issue 
directly, but do outline the essentials of dynamical encounters of that kind,
in order to facilitate future discussions.

Basics of the formalism we outline have been in the literature for
some time. Our objective  is to set them down as a package, extending them as needed for
application to the issues at hand, while looking for
parameterizations that may be useful for diagnostic purposes. We 
specifically target AGN outflow/shock encounters, although the necessary formalism has broader applications.
We also include along the way some preliminary results from
simulations in order to illustrate the kinds of behaviors that result.
We ignore here a number of complications that will ultimately be important to
detailed models, but that do not seem essential to a ``first order'' picture.

Although our goal in this paper is to set out a simple set of
analytic relationships, we do show illustrations of behaviors from high resolution
numerical simulations that will be discussed in detail in subsequent publications.
Those simulations were carried out using the
WOMBAT MHD code employing a second order MHDTVD Riemann solver \cite{men17}.
An adiabatic equation of state with $\gamma = 5/3$ was employed.
Steady bi-polar jets were formed within a small cylindrical volume, where the jet
pressure, $P_j$, was kept approximately equal to the ambient pressure, and the
jet plasma was accelerated to a velocity, $v_j$. All the simulations
shown in this paper kept the density within the jet-source cylinder, $\rho_j$, at 1\% of the undisturbed
ambient density.
We have, however, varied these parameters to confirm relations described below.
We also allowed jet power to cycle off in some cases.
The simulated jets were weakly magnetized with a  toroidal geometry
and nominal $\beta = P_g/P_m = 75$ within the jet-source cylinder. Magnetic
stresses are too small to influence dynamics at the levels relevant to this paper. The
simulations did include CRe; in follow-up
discussions we will discuss in detail synchrotron emission properties. That is beyond
the scope of the present paper, however.

The paper outline is this. Section \ref{sec:shock-cavity} reviews the interactions between a plane shock
and a low density cavity. In section \ref{sec:jet-wind} we outline in simple terms  the propagation of
a jet into a head or tail wind (section \ref{subsec:term}) and into a cross wind (section \ref{subsec:cross}) that would, for instance, represent post-shock ambient conditions. For completeness section \ref{sec:cocoon} 
sets out a simple model to estimate the form of the plasma cocoon inflated by jets nominally prior to
the above interactions.

\section{An ICM Shock Colliding with a Cavity}
\label{sec:shock-cavity}

Typically the initial encounter between an ICM shock and an AGN will involve the
shock penetrating the AGN cocoon/cavity inflated by the AGN outflow.
The basic dynamics of such an encounter are qualitatively simple: the
core of the cavity is crushed relatively quickly, while the perimeters of the cavity roll into
a vortex ring structure \cite{eb02}. Here we simply outline of the
physics.

Consider then a planar ICM shock of Mach number, $\mathcal{M}_i = v_{si}/a_i$, where $v_{si}$ is the
speed of the shock in the ICM and $a_i = \sqrt{\gamma_i P_i/\rho_i}$ is the ICM sound
speed. $P_i$ and $\rho_i$ are the pre-shock ICM pressure and mass density, while $\gamma_i$ ($=5/3$) is the ICM adiabatic index.
Typical ICM shocks will have modest Mach numbers, $\mathcal{M}_i \la 4$, \cite{rkhj03} but to
allow a wider application of this discussion we do not restrict $\mathcal{M}_i$.
We then assume the shock encounters a low density cavity, which we take
initially to be in pressure
equilibrium with the ICM ($P_c = P_i$). Let the undisturbed density inside the cavity be $\rho_c = \delta \rho_i$, with $\delta \ll 1$.
The initial cavity  sound speed is, $a_c = \sqrt{\gamma_c/\gamma_i}~a_i/\sqrt{\delta} \sim a_i/\sqrt{\delta} \gg a_i$, where, acknowledging that
the cavity includes relativistic plasma, we allow the cavity
potentially to have an adiabatic index, $\gamma_c \le \gamma_i$, distinct from the ICM.

As the shock encounters the cavity, it penetrates at a speed $v_{sc} = \mathcal{M}_c a_c \ge a_c$
pulling ICM gas with it. Pressure balance between the original cavity and
un-shocked ICM leads to the condition, $v_{sc} \ge v_{si}/(\delta^{1/2} \mathcal{M}_i)$.
The original ICM-cavity contact discontinuity (CD) follows
the shock at a speed, $v_{cD}$, intermediate between the ICM shock and the cavity shock.
At the same time a rarefaction propagates back into the post-shock ICM at
the post-shock ICM sound speed.

Pfrommer and Jones \cite{pj11} presented the full, nonlinear, 1D Riemann solution to this problem. Here
we need only a few, approximate behaviors in order to understand what happens during
the  encounter. Generally,
even though the shock speed is greater inside the cavity, the shock strength is less than the incident shock; that is, $\mathcal{M}_c < \mathcal{M}_i$. There is no
general, analytic formula for $\mathcal{M}_c$, although it can be found numerically from
the Riemann solution.  In the strong shock limit, when $\mathcal{M}_c \gg 1$, Pfrommer and Jones found $v_{sc} \sim 2~v_{si}$; that is
$\mathcal{M}_c \sim 2~ \delta^{1/2} \mathcal{M}_i$. In the weak shock limit (as $\mathcal{M}_c \rightarrow 1$), of course,
$v_{sc} \sim v_{si} \delta^{-1/2}$. In either regime the internal shock speed
considerably exceeds the incident, ICM shock speed whenever $\delta \ll 1$. 
Consequently, the shock passes through the cavity much more quickly than it
propagates around it. 

The velocity of
the CD within the cavity, $v_{CD}$, obtained in the Riemann solution is,
\begin{equation}
v_{CD} = \frac{2}{\gamma_c+1}\frac{(\mathcal{M}_c^2 - 1)}{\mathcal{M}_c^2}v_{sc}.
\end{equation}
This measures the cavity collapse rate. It will generally
be less than $v_{sc}$. However, Pfrommer and Jones found, so long as $\mathcal{M}_i \ga 2$  and $\delta \ll 1$, that
$v_{CD} > v_{si}$. Under those circumstances, the cavity is crushed (the initial CD pushes all the way through the cavity) {\it before} the
ICM shock propagates around it.

We note that
if the cavity has previously developed a thick boundary layer within which $\delta$ is not very small (e.g., through turbulent mixing; see section \ref{sec:cocoon}), the internal
shock speed within this layer will be slower than above, but $\mathcal{M}_c$ can remain comparable to $\mathcal{M}_i$ through that
boundary layer. This can, for instance, significantly influence such things as
particle acceleration during shock passage and shock amplification of turbulence.

If the initial cavity is ``round'' (e.g., a sphere), the shock intrusion into the cavity begins sooner
and is usually stronger at the normal,  ``point of first contact.'' On the other hand, the oblique penetration towards the
extremes of the cavity boundary also generates vorticity. An initially spherical
cavity will, thus, evolve into a torus, or vortex ring that mixes ICM and cavity plasma \cite{eb02,pj11}. 
 Analogously, even if the cavity boundary is planar, but the shock normal is oblique to
the cavity boundary, refraction of the shock and penetrating CD will produce vorticity
and mixing \cite{pjr15}. Examples of shocks impacting ICM cavities
have been presented previously \cite{eb02,pj11}. Outcomes are also evident
in Figures \ref{fig:aljet} and \ref{fig:orjet}.

\section{Jet Propagation in a (Post-shock) Wind}
\label{sec:jet-wind}

\begin{figure}
\includegraphics[width=0.48\textwidth,height=0.41\textwidth]{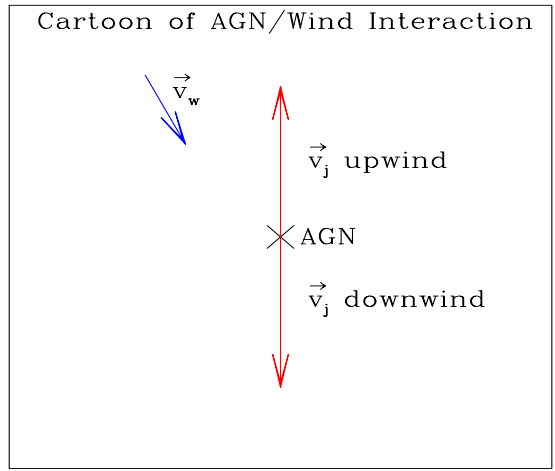}
\caption{\label{fig:cartoon} Cartoon illustrating the basic geometry of bipolar AGN jets
propagating into an ambient wind.}
\end{figure}

If an AGN jet remains active following shock impact, so that it drives into the post-shock ICM, its
propagation will be modified because the post-shock ICM is denser than the
pre-shock ICM and also because the post-shock ICM plasma is put in motion by the shock. 
In this section we outline a toy model to address these effects. The basic geometry is illustrated in Figure \ref{fig:cartoon}.
We
approximate the dynamics as that of a jet within an ambient wind, decomposing the
wind velocity into components parallel (head wind or tail wind) and orthogonal (cross wind) to the
jet velocity vector, $\vec{v_j} = v_j\hat{v_j}$, near its terminus. That is, we set
\begin{equation}
\vec{v}_w = v_{w,\parallel}\hat{v}_j + v_{w,\perp} \hat{v}_\perp,
\end{equation}
where $\hat{v}_j\cdot \hat{v}_\perp = 0$.
When $v_{w,\parallel} \ne 0$; that is, whenever there is a tail wind, $v_{w,\parallel}> 0$, or a head wind, $v_{w,\parallel} < 0$, the rate of advance of the jet terminus will be modified; 
when $v_{w,\perp} \ne 0$ the jet propagation will be deflected laterally.
Although we present the model in the context of a post-shock flow, the formalism
applies to any relative motion between the AGN and its ambient medium. Indeed the model
is essentially a merger and modest extension of classical cartoons of AGN jet propagation \cite{bbr84,brb79}.
Simply stated, jet advance and deflection are influenced by parallel and
transverse ram pressures created by the wind.

\subsection{Advance of a Jet Terminus in an Ambient Head or Tail Wind}
\label{subsec:term}

We begin by considering the influence of an ambient wind motion aligned with the jet velocity; that is,
consequences of $v_{w,\parallel} \ne 0$.
To keep the discussion simple we consider here only steady, collimated
jets (zero opening angle) and homogeneous ambient media. We adopt the standard picture that the advance of the jet
terminus into the ICM can be expressed in terms of the propagation of
a contact discontinuity that forms at the head of the jet \cite{bbr84}. {\it Relative to the
AGN}, that boundary (the jet ``head'') propagates at velocity, $v_h\hat{v}_j = d \ell_j/ dt~\hat{v}_j$, where $\ell_j$ represents the length of the jet from the AGN. There will
be a bow shock propagating into the wind, ahead of the head, as well as a
reverse, ``terminal'' shock in the jet. The simple, 1D cartoon model ignores these
details, drawing a box around all this and assumes the jet thrust; that is, the total jet momentum flux coming
in through an area, $A_j = \pi r_j^2$, is balanced by the total wind momentum flux
coming in on the opposite side of the box, again, through an area, $A_j$.
In reality, because both jet and wind plasma will, as ``back flow'', exit the face of
the box through which the jet enters (and box sides), the effective sizes of the box faces need to be
larger than $A_j$. We can crudely account for this asymmetry by
setting the effective area allocated to the wind, $A_h$, to
be larger than the nominal cross section of the jet.

We comment, as well, that even for a collimated jet the radius, $r_j$, will not generally be a
constant along the length of the jet, especially when it becomes over or under pressured
with respect to its surroundings. In that case, for instance, the propagating jet plasma will execute a sequence of
expansions and rarefactions around an equilibrium radius. Consequently,
the jet velocity, $v_j$, density, $\rho_j$, and pressure, $P_j$, will all vary along
a real jet. Our analysis here does not
try to account for dynamics at that level of detail, so we will assume below
that the jet has some suitably chosen characteristic radius, $r_j$, density, $\rho_j$ and pressure, $P_j$.

We allow both the jet velocity and equation of state (EoS) to be either non-relativistic or relativistic, 
although we will assume the propagation of the wind and the head through the wind in the
AGN reference frame are non-relativistic.
Then, the jet 4-velocity is $U_j = \Gamma_j v_j$, with $\Gamma_j$ the Lorentz factor
of the jet velocity, $v_j$, while the enthalpy density of
the jet plasma in the jet plasma frame is $w_j c^2 = e_j  + P_j = \rho_j c^2 + e_j' + P_j = \rho_j c^2 \tilde{w}$, 
with $e_j$ the internal energy density of the jet plasma, including rest mass energy, $\rho_j c^2$. 
The jet momentum flux density in the AGN frame can be written \cite{llfm}
\begin{equation}
T_{mj} = w_j U_j^2 + P_j = \rho_j \tilde{w} U_j^2 + P_j.
\end{equation}
The jet thrust is 
\begin{equation}
N_j = T_{mj} A_j = (w_j U_j^2 + P_j) A_j = \rho_j U_j^2 ( \tilde{w}_j+ \frac{P_j}{\rho_j U_j^2} ) A_j.
\end{equation}
For a non-relativistic EoS, where $e \approx \rho c^2$, $\tilde{w} = (e + P)/(\rho c^2)  \approx 1 + (e'+P)/(\rho c^2) = 1 + a^2/((\gamma - 1) c^2) \approx 1$, while for a relativistic EoS, where $P \approx (1/3) e$, $\tilde{w} \sim 4 P/(\rho c^2) \gg 1$. 

It will be useful later on also to have expressions for the jet energy flux density in the
AGN frame. Subtracting off the rest mass energy this is
\begin{eqnarray}
T_{ej} = U_j \Gamma_j (w c^2 - \rho_j c^2/\Gamma_j)\nonumber\\
= U_j \Gamma_j \rho_j c^2 (\tilde{w} - 1/\Gamma_j)\nonumber\\
= \Gamma_j^2 \rho_j c^2 v_j \left[(1 -  1/\Gamma_j) + (e_j' + P_j)/(\rho_j c^2)\right],
\end{eqnarray}
so that the ``luminosity'' of the jet is
\begin{equation}
L_j = T_{ej} A_j = v_j \Gamma_j^2 \rho_j c^2 (\tilde{w} -  1/\Gamma_j) A_j.
\label{eq:jetlum}
\end{equation}

We define the {\it internal} jet Mach number as, $\mathcal{M}_j = \Gamma_j v_j /(\Gamma_{s,j} a_j)$,\cite{k80} 
where $a_j = c \sqrt{\partial P_j/\partial e_j}$ is the jet sound speed, with 
$\Gamma_{s,j}^2 = 1/(1 - (a_j/c)^2)$. 
For a relativistic EoS $a^2 = (1/3) c^2$, while $\Gamma_s^2 = 3/2$, so $\Gamma_s^2 a^2 = (1/2) c^2$. 
Note in this limit that the jet Mach number, $\mathcal{M}_j \to \sqrt{2}~\Gamma_j (v_j/c)$, 
which, if $v_j \to c$ gives $\mathcal{M}_j = \sqrt{2}~\Gamma_j$. This makes clear
that the fact that a jet velocity is close to the speed of light does not,
by itself, imply a high Mach number for the jet dynamics.
A non-relativistic EoS with $P = (\gamma - 1) e'$ recovers the familiar $a^2 = \gamma P/\rho$.

With these definitions the momentum flux density is  
\begin{equation}
T_{mj} = \rho_j \tilde{w} \Gamma_j^2 v_j^2 + P_j = \mathcal{M}_j^2 P_j \left[\frac{a_j^2}{c^2 - a_j^2} \tilde{w}\frac{\rho_j c^2}{P_j} + \frac{1}{\mathcal{M}_j^2}\right].
\end{equation}
Setting  $\gamma' = (\Gamma_{s,j}^2 a_j^2 w_j)/P_j$, this becomes simply
\begin{equation}
T_{mj} = \gamma' \mathcal{M}_j^2 P_j \left[1 + \frac{1}{\gamma' \mathcal{M}_j^2}\right].
\label{eq:momflux}
\end{equation}
For a relativistic EoS $\gamma' = 2$, whereas for a non-relativistic EoS, $\gamma' = \gamma$, the usual gas adiabatic index.
Written in this form the expression for the jet momentum flux is virtually the
same for relativistic or non-relativistic jets.
If, as is often the case, $\mathcal{M}_j^2 \gg 1$ we can neglect the second term in square brackets.
Then, the jet thrust for either relativistic or non-relativistic flow is
\begin{equation}
N_j = T_{mj} A_j \approx \gamma' \mathcal{M}_j^2 P_j A_j.
\label{eq:momflux-ap}
\end{equation}
In either regime the thrust depends simply on the internal jet pressure and
the internal jet Mach number.

The jet luminosity is not quite so tidy, but still simple to express.
For a non-relativistic EoS 
the luminosity would be
\begin{equation}
L_j  \approx \mathcal{M}_j \Gamma_j \rho_j c^2~(\Gamma_{s,j}a_j)~\left[1  + \frac{a_j^2}{(\gamma_j - 1) c^2} - \frac{1}{\Gamma_j}\right] A_j,
\end{equation}
which, with  $v_j \ll c$, but $v_j/a_j = \mathcal{M}_j \gg 1$ and $\gamma_j = 5/3$ takes the simple familiar form,
\begin{equation}
L_j \approx \frac{1}{2} \rho_j v_j^3 A_j = \frac{5}{6} \mathcal{M}_j^2 v_j P_j A_j.
\label{eq:ljet-nonrel}
\end{equation}
At the other extreme, with a relativistic EoS, $\tilde{w}_j = 4 P_j/(\rho_j c^2)$, and $\Gamma_j \gg 1$, the luminosity is also simple to express; namely,
\begin{equation}
L_j \approx 2 \mathcal{M}_j^2 c P_j A_j,
\label{eq:ljet-rel}
\end{equation}
since, now $2 \Gamma_j^2 \approx \mathcal{M}_j^2$.
Again, in this form the relativistic and non-relativistic expressions are almost the
same. 
The jet luminosity can, like the thrust, be described in terms of the jet pressure
and Mach number, but now also in terms of a jet speed, $v_j$, (which may $\to c$).


We return now to estimating the rate at which the jet terminus propagates through its
surrounding plasma.
The momentum flux balance condition determining the propagation velocity, $\vec{v}_h$, is actually measured in the frame of the head, so we
should transform the momentum flux relations to that frame. 
However, provided $v_h/v_j \ll (c/v_j)^2/\Gamma_j^2$, it is easy to show that the fractional change in $T_{mj}$ is small. We will neglect that correction below.
The momentum balance condition becomes,
\begin{equation}
N_j = A_j T_{mj} = A_h [\rho_w (v_h - v_{w,\parallel})^2 + P_w ].
\end{equation}
Using equation \ref{eq:momflux-ap} this leads to
\begin{equation}
\frac{(v_h - v_{w,\parallel})^2}{a_w^2}\left[ 1 + \frac{a_w^2}{\gamma_w (v_h \mp v_w)^2}\right] \approx 
 \mathcal{M}_j^2 \frac{\gamma'}{\gamma_w}\frac{A_j P_j}{A_h P_w} 
 \label{eq:headv-vel}
\end{equation}
or,
\begin{equation}
(\mathcal{M}_h \mp \mathcal{M}_{w,\parallel})^2\left[ 1 + \frac{1}{\gamma_w (\mathcal{M}_h \mp \mathcal{M}_w)^2}\right] \approx 
 \mathcal{M}_j^2 \frac{\gamma'}{\gamma_w}\frac{A_j}{A_h}\frac{P_j}{P_w}
 \label{eq:headv-mach}
\end{equation}
where $\mathcal{M}_h = v_h/a_w$ and $\mathcal{M}_{w,\parallel} = |v_{w,\parallel}|$
are the Mach numbers of the jet head advance and the head/tail wind with respect to the AGN. Then, of course, $a_w = \sqrt{\gamma_w P_w/\rho_w}$ is the wind sound speed.
In equations \ref{eq:headv-vel} and  \ref{eq:headv-mach} the upper (lower) sign corresponds
to a tail (head) wind.

If $(\mathcal{M}_h \mp \mathcal{M}_{w,\parallel})^2 \gg 1$ we have the simple result
\begin{equation}
|\mathcal{M}_h \mp \mathcal{M}_{w,\parallel}| \approx \mathcal{M}_j \sqrt{\frac{A_j}{A_h}}\sqrt{\frac{P_j}{P_w}}\sqrt{\frac{\gamma'}{\gamma_w}}.
\label{eq:headv-app}
\end{equation}
That is, the Mach number of the advance of the head relative to its ambient medium
is similar to the internal Mach number of the jet modified by  a factor that depends
mostly on  the ratio of the integrated jet pressure, $ P_j A_j$, to the
integrated wind pressure (isotropic pressure, not ram pressure) ``across the head'', 
$P_w A_h$. In our simulation experiments with steady, fixed axis, non-relativistic jets $A_j/A_h \sim 1/2$ provides
reasonable matches between simulated jet propagation and equations \ref{eq:headv-vel}, \ref{eq:headv-mach} and \ref{eq:headv-app}. 

Some insights into appropriate $P_j/P_w$ are useful going forward.
Our discussions relate to jets that are pressure confined as they propagate. So, we
expect $P_j \sim P_{a}$, where $P_{a}$ represents the pressure of whatever
medium is immediately surrounding the jet. If a jet enters a region where it is out of
pressure balance, it will generally expand or converge to compensate. As noted
above, the pressures within simulated jets are actually quite non-uniform as a result. On average, however, we find, independent of the pressure assignd to a simulated jet at its
origin, the average pressure along the propagating jet becomes roughly comparable to the ambient pressure. We also find in simulations and argue in section \ref{sec:cocoon} that the pressure inside jet cocoons
are commonly roughly similar to those in the undisturbed surroundings, even though
the cocoon formation drives a shock into its surroundings. Consequently,
we expect within a factor of a few that $P_j/P_w \sim 1$. Nonetheless, we leave such ratios as undetermined in our
analyses, in order to reveal their roles.

We note also for non-relativistic jets with non-relativistic EoS that, since  for 
both the jet and the ambient medium, $a^2 = \gamma P/\rho$, equation \ref{eq:headv-app} can be written
\begin{equation}
|v_h \mp |v_{w,\parallel}||  \approx v_j \frac{\sqrt{A_j/A_h}}{\sqrt{\rho_w/\rho_j}},
\end{equation}
independent of $P_j/P_w$. This recovers the commonly applied assumption that, modulo ``an efficiency factor'' ($\sqrt{A_j/A_h}$) the advance speed of the jet head is roughly the
jet speed reduced by a factor of the square root of the density ratio between the
ambient medium and the jet when the jet and its advance are both highly supersonic \cite{bbr84}.

\begin{figure}
\includegraphics[width=0.48\textwidth,height=0.41\textwidth]{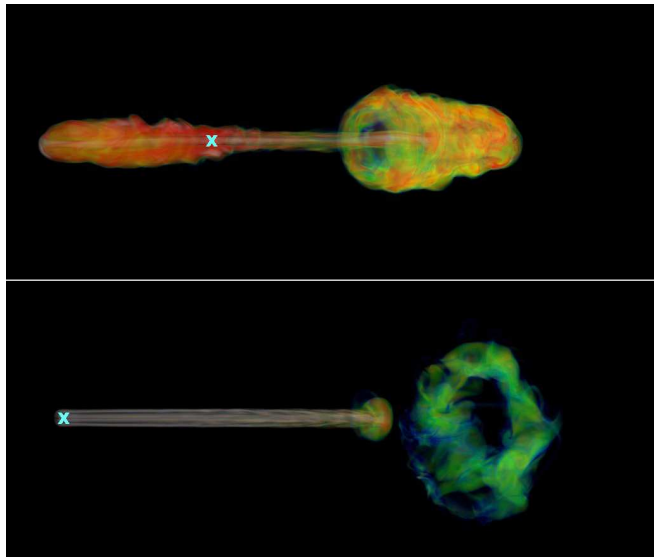}
\caption{\label{fig:aljet}Illustration from simulations of ICM shocks colliding head on with $\mathcal{M}_j = 3.5$
AGN jet pairs. Both panels are volume renderings of a passive jet mass-fraction tracer. Shock
propagation was left to right. 'X' marks AGN location. The jet axes are rotated 40 degrees out of the plane of the sky with the left jets approaching the observer. {\bf Top}: $\mathcal{M}i = 2$. {\bf Bottom}: $\mathcal{M}_i = 4$. See text for details.}
\end{figure}

It is further evident from these relations, as we would expect intuitively, that the advance speed of the jet with respect to
the AGN, $v_h$, is greater when it propagates downstream with a wind (a ``tail wind'')
than if it propagates upstream into a wind (a ``head wind'').
In fact a jet propagating into a sufficiently strong head wind can be stopped,
or even reversed. Using equation \ref{eq:headv-app} the approximate condition for
the head wind to stop forward progress of the head is
\begin{equation}
|\mathcal{M}_{w,\parallel}| \ga \mathcal{M}_j \sqrt{\frac{A_j }{A_h}}\sqrt{\frac{\gamma' P_j}{\gamma_w P_w}}.
\label{eq:headstop}
\end{equation}
We have verified this condition in our simulations (see Figure \ref{fig:aljet}).

Although the above results would be applicable for any AGN relative motion through
its ambient medium when the AGN jets are aligned with the relative motion, we introduced
the issue in the context of post-shock ICM flows. In that case it is useful to express
these relations in terms of wind properties resulting from ICM shocks. To keep it
simple, we assume that the pre-shock ICM plasma was at rest with respect to the AGN, although that is easily modified. Again expressing
the ICM shock Mach number as $\mathcal{M}_i$ we have (assuming $\gamma_w = \gamma_i = 5/3$),
\begin{eqnarray} 
\label{eq:jump-d}
\rho_w = \frac{4\mathcal{M}_i^2}{\mathcal{M}_i^2+3}\rho_i,\\
\label{eq:jump-p}
P_w = \frac{5\mathcal{M}_i^2-1}{4}P_i,\\
\label{eq:jump-v}
v_w = \frac{3}{4}\frac{\mathcal{M}_i^2-1}{\mathcal{M}_i}a_{i},\\
\label{eq:jump-a}
a_w^2 = \frac{(\mathcal{M}_i^2 + 3)(5 \mathcal{M}_i^2 - 1)}{16 \mathcal{M}_i^2} a_i^2.
\end{eqnarray}
where, as above, the subscript index `i' identifies the unshocked ICM, and `w'
indicates post-shock ICM (the ``wind'').

Applying equations \ref{eq:jump-d} - \ref{eq:jump-a} to equation \ref{eq:headstop} we obtain an approximate relation
for the strength, $\mathcal{M}_{is}$, of an ICM shock that can stop the advance of an 
approaching jet of internal Mach number, $\mathcal{M}_j$ in a 
``head on collision'' when the AGN is at rest in the undisturbed ICM; namely
\begin{equation}
\mathcal{M}_i =\mathcal{M}_{i,s} ~\ga~\frac{2}{3} \sqrt{\frac{3\gamma'}{5}} \mathcal{M}_j \sqrt{\frac{P_j}{Pi}}\sqrt{\frac{A_j}{A_i}}\frac{1}{G(\mathcal{M}_i)},
\end{equation}
where $G(\mathcal{M}_i) = (1 - 1/\mathcal{M}_i^2)/\sqrt{1 + 3/\mathcal{M}_i^2}) \le 1$. As noted above, 
we expect for steady jets with fixed axes, $A_j/A_i \sim 1/2$. The function  $G(\mathcal{M}_i) \sim 0.6 - 0.9$,  
for $2 \la \mathcal{M}_i \la 4$,
corresponding to expected ``merger-related''  ICM shock strengths.
As a rough rule of thumb, then,
an AGN jet running head on into a shock will be stopped or reversed if the
Mach number of the shock is comparable  to or even a bit less than the Mach number of the jet.

The above behaviors are illustrated in Figure \ref{fig:aljet} for two, simulated
non-relativistic 
shock-jet interactions with different relative shock strengths. The views are
3D volume renderings of jet mass fraction; that is, only plasma that originated
from the AGN is shown. In both cases, 
$\mathcal{M}_j = 3.5$, and the shock normal was aligned with the (bi-polar) jet axis.
ICM shock propagation was left-to-right, although in this view the shock plane  was rotated 40 degrees
from the line of sight, in order to reveal the structures more clearly.
The location of the AGN is marked by an X in each case. The color map of the mass fraction tracer runs from ``white'' (100\%) through yellow, red, green and blue ($\sim$ 30\%).

The figure upper panel represents the outcome for an ICM shock with $\mathcal{M}_i = 2 = 0.57 \mathcal{M}_j$, 
whereas the lower panel involves a $\mathcal{M}_i = 4 = 1.1 \mathcal{M}_j$ shock. Both
are shown at approximately the same time interval since initial contact between the
shock and the left-facing  (upwind) jet. In the bottom panel the stronger shock has left the computational domain
to the right after reversing the left-facing jet and crushing the two original jet plasma cocoons and stripping them from the jets. The ``smoke ring'' to the right is the resulting vortex ring structure, whose
formation  out of the pre-shock cavities was outlined in the previous section. 
The flaring seen at the right end of the remaining (downwind) jet represents the head of that
jet. The shocked jet plasma is unable to propagate back to the
AGN through the strong right-facing wind behind this shock. It is not able to refill
a cocoon (see section \ref{sec:cocoon}). The jet remains ``naked''.
Several AGNs in disturbed clusters have been seen that are candidates for this 
interaction. Probably the best known example is the radio source `C' in A2256, \cite{rsm94,or14} which has a roughly 1 Mpc,
very thin (unresolved) ``tail'' extending to the west of an AGN, but nothing evident to
the east. Indeed, that AGN is seen projected near a strong ``radio relic`` in the cluster, suggesting
that it could have passed through a moderately strong cluster merger shock.

The weaker and slower ICM shock involved in the upper Figure \ref{fig:aljet} panel dynamics is still in the volume illustrated, although
not directly visible in this rendering. In the undisturbed
ICM its position along the jet axis is about 2/3 of the distance from the AGN X to the right end of the
right-facing jet head. Inside the right-side cocoon, the shock has just reached the
head at this time. We found experimentally in this case that an incident shock
with $\mathcal{M}_i \approx (3/4)  \mathcal{M}_j$ would just stop the approaching (left facing in the figure) jet.

Thus, evidence for relative AGN foreshortening to one side has the potential to find and even measure the strengths of ICM shocks or winds.

\subsection{Deflection of a Jet by a Cross Wind}
\label{subsec:cross}

\begin{figure}
\includegraphics[width=0.48\textwidth,height=0.41\textwidth]{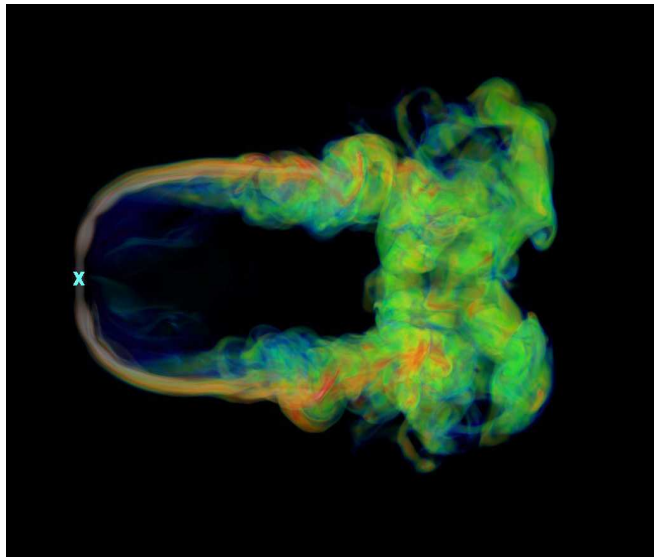}
\caption{\label{fig:orjet}Simulation of an $\mathcal{M}_i = 4$ shock that crossed from
the left and collided at right
angles with an $\mathcal{M}_j = 10$
AGN jet pair (vertical jet axis in the image). Shown is a volume rendering of a passive jet mass-fraction tracer
after the shock has passed through the volume shown. 
'X' marks AGN location. View orientation is the same as Figure \ref{fig:aljet}. See text for details.}
\end{figure}

In the presence of a cross wind, $v_{w,\perp} \ne 0$ (see Figure \ref{fig:cartoon}), the jet is subjected to an unbalanced transverse
ram pressure force, $\rho_w v_{w,\perp}^2$. Particularly, if that cross wind results
from a crossing shock, the jet cocoon (cavity) will be crushed and stripped away from
the propagating jet (see Figure \ref{fig:orjet}). From that point on the jet interacts
directly with the wind, and we can estimate the induced transverse pressure gradient within
the jet as 
\begin{equation}
\partial P_j/\partial x_{\perp} \approx \rho_w v_{w,\perp}^2/(2 r_j).
\label{eq:pgrad}
\end{equation}
Then the transverse acceleration of a steady jet is determined by the relation
\begin{equation}
\frac{w \Gamma_j^2 v_j}{c^2}\frac{\partial v_{j,\perp}}{\partial \ell} \approx \frac{\rho_w v_{w,\perp}^2}{2 r_j}(1 - \Gamma_j^2 v_{j,\perp}^2/c^2),
\label{eq:ftrans}
\end{equation}

We can use $\partial v_{j,\perp}/\partial\ell \sim v_j/\ell_b$ as a way to estimate
the length $\ell_b$ over which the transverse ram pressure from the wind will
deflect the jet by 90 degrees; that is, the `` jet bending length''. Neglecting the (initially small)  second term inside the parentheses in
equation \ref{eq:ftrans} this gives,
\begin{equation}
\ell_b \approx 2 r_j \frac{w \Gamma_j^2 v_j^2}{\rho_w v_{w,\perp}^2} \approx 2 r_j \frac{T_{mj}}{\rho_w v_{w,\perp}^2},
\end{equation}
where the final expression has used equation \ref{eq:momflux-ap}.
Analogous to the results of the previous subsection we can also write this
final form in terms of the jet Mach number, $\mathcal{M}_j$ and the cross-wind Mach number, $\mathcal{M}_{w,\perp}$,
\begin{equation}
\ell_b \approx 2 r_j \frac{\gamma'}{\gamma_w}\frac{\mathcal{M}_j^2}{\mathcal{M}_{w,\perp}^2}\frac{P_j}{P_w}.
\label{eq:ellbend}
\end{equation}
If the jet and wind pressures are comparable, the ratio of the jet bending length, $\ell_b$,
to the jet diameter, $2 r_j$, is roughly $(\mathcal{M}_j/\mathcal{M}_w)^2$.
A little bit of algebra shows that equation \ref{eq:ellbend} matches the bending radius of curvature derived in \cite{brb79} for
an AGN moving supersonically through an ambient medium at right angles to the jet axis. 
Of course, only relative motion matters, and here we emphasize that it is not necessary that the
cross wind is supersonic for the bending to develop. The bending radius will, however,
scale inversely with the square of the Mach number of relative motion. So, once again,
obvious distortion in the AGN points to relatively large Mach number of the
relative motion between the AGN and its immediate surroundings.

Assuming the cross wind under discussion comes entirely from an ICM shock propagating transverse to the
jet axis, we can use equations \ref{eq:jump-d} - \ref{eq:jump-a} to give a relation between the jet bending length,
the jet radius, the jet Mach number and the ICM shock Mach number, $\mathcal{M}_i$, 
\begin{equation}
\ell_b \approx \frac{8}{9}\frac{\gamma'}{\gamma_i}\frac{\mathcal{M}_j^2 (\mathcal{M}_i^2 + 3)}{(\mathcal{M}_i^2 - 1)^2}\frac{P_j}{P_i}~ r_j.
\label{eq:lbend-mi}
\end{equation}

By convention, when the jet length, $\ell_j$ satisfies $\ell_j~\ga~\ell_b$, but $\ell_b\gg r_j$ the resulting
AGN morphology would be describe as a ``wide angle tail'' radio galaxy, or a ``WAT". As
the ratio $\ell_b/r_j$ becomes smaller, the morphological label would shift to ``narrow angle tail'' or ``NAT''. 

Figure \ref{fig:orjet} illustrates results from a simulation of a $\mathcal{M}_i = 4$ shock that crossed from the left and collided at right angles with a pair of $\mathcal{M}_j = 10$ jets oriented
vertically in the image. 
Equation \ref{eq:lbend-mi} predicts
$\ell_b/r_j = 6.7$, which is actually very close to the empirical result of the
simulation. 
As in Figure \ref{fig:aljet} this image shows
a volume rendering of the passive jet mass fraction tracer.
The shock plane has again been rotated 40 degrees from the line of sight to
make 3D structures more distinct.  We comment on the close resemblance between 
the structures visible in Figure \ref{fig:orjet} and the ``classic'' NAT  source, NGC1265 \cite{odea86}, whose morphology has long been modeled in terms of a strong cross wind in
the rest frame of the host galaxy. Here the wind is a feature of the post-shock environment. That has also been suggested for
NGC1265 \cite{pj11}.

Note that the two jets  in Figure \ref{fig:orjet} remain stable long after they are deflected by $~\sim~90$ degrees into ``tails'', 
even though turbulent mixing regions develop around them and enlarge downstream.
The tails merge at the far right into a pair of merging vortex rings that
formed along the lines outlined in section \ref{sec:shock-cavity} as the ICM shock crushed the two jet cocoons produced before shock impact.
The jets can be traced as coherent structures almost to the vortex rings.
The shock itself has left the observed volume to the right at the time shown.

Finally, we comment on the more complex case where an incident shock collides at an arbitrary angle with respect
to the AGN jet axis, or more generally when the AGN jets encounter an oblique wind. Since $v_{w,\perp}$ is initially the same for both jets, the rate of jet deflection
indicated by equation \ref{eq:ftrans} would, before deformation, be the same for both jets (see Figure \ref{fig:cartoon}). 

The aligned wind velocity,
$v_{w,\parallel}$, has the same magnitude, but opposite sign on the two
jets. The downwind jet, where $v_{w,\parallel} > 0$, thus advances more rapidly.
As the jets begin to be deflected, however, the upwind jet, where $v_{w,\parallel} < 0$, is deflected to be more nearly transverse
to the wind ($v_{w,\perp}$ increases along the jet on that side), while the downwind jet is deflected to be more nearly aligned with the wind ($v_{w,\parallel}$ increases along this jet). 
Consequently, the  upwind jet becomes more sharply bent 
than the downwind jet. We have confirmed in simulations a regular transitioun along these lines between the aligned wind interactions described in section \ref{subsec:term}. and cross wind interactions outlined at the beginning of this section.  Evidently, such asymmetries can provide information
about both the relative Mach numbers of the jets and a wind and the relative orientations
of the wind and the jets. The shape evolution of the two jets depends distinctively
on the ratio of two Mach numbers as well as the orientation between the AGN jet axis
and the wind velocity vector, $\vec{v}w$. Thus, the resulting shape provides a
means to determine both $\mathcal{M}_j/\mathcal{M}_w$ and $\hat{v}_w\cdot \hat{v}_j$.
Also, it is obvious that if the ``weather conditions'' encountered by jets vary
as they extend through the ICM, additional morphological features are likely to develop that can
be used to identify and characterize these ICM structures.

\section{Formation of a Backflow Cocoon}
\label{sec:cocoon}

We began this discussion with an outline of shock propagation through a  pre-existing
cavity formed
by AGN outflows; that is a jet cocoon. Then we outlined propagation of the jets themselves in the post-shock flow. 
The shocks both crush the cocoon and may, if the wind ``stripping'' is faster
than replacement from the jet terminus, remove the cocoon (see Figures \ref{fig:aljet} and \ref{fig:orjet}).
We did not, however, address what might be learned from the cocoons themselves,
In order to tie the pieces together, here we outline briefly some of the related physics
connecting the jet propagation to the formation of the jet cocoon before
shock impact.  Since our purpose is only to lay out a rough picture of cocoon inflation, we ignore
in this section complications such as relative motions between the AGN and its ambient ICM,
the presence of large scale pressure or density gradients, or ICM turbulence. All of these
will modify cocoon morphology and quantitative measures, although they should not fundamentally
change the basic picture presented here. 

In simple terms the cocoon represents the reservoir of plasma that has previously passed through
the jet. The cocoon plasma will generally be of substantially lower density
than the surrounding ICM, with the two media nominally separated by a contact discontinuity.
Because it is also a ``slip surface'', that boundary is likely to be unstable to Kelvin-Helmholtz instabilities (KHI), so that some
degree of mixing will take place (unless, for instance, magnetic fields in or around the
cocoon are strong enough and coherent enough to stabilize the contact discontinuity \cite{rebhp07,odj09}). 

\begin{figure}
\includegraphics[width=0.48\textwidth,height=0.41\textwidth]{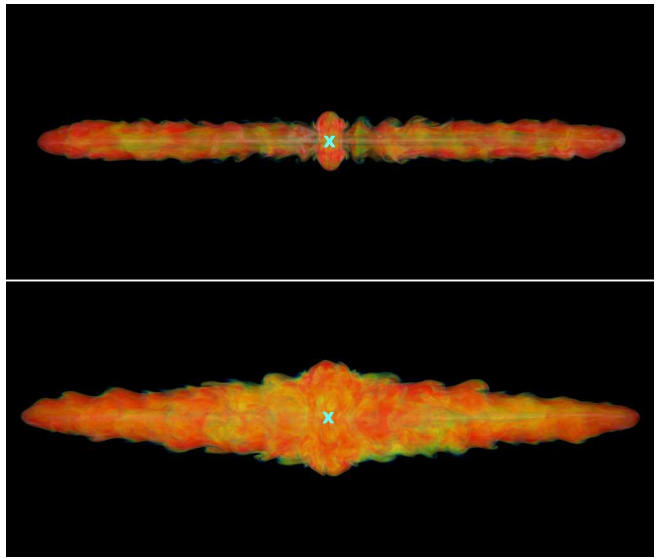}
\caption{\label{fig:noshock-color} Jet ``backflow'' cocoons formed by $\mathcal{M}_j = 3.5$ ({\bf{Top}}) and $\mathcal{M}_j = 10$ ({\bf{Bottom}}) steady jets. A passive, jet mass-fraction
tracer is volumed rendered in the images. The jets are ``in the plane of the sky''
and shown when they have approximately the same lengths, $\ell_j$.
The location of the AGN is marked by an `X'. See text for details.}
\end{figure}

Figure \ref{fig:noshock-color} shows volume renderings of cocoons formed from two
simulated steady AGN jet pairs. In each case the AGN (marked by an `X') expelled identical, but oppositely directed jets. 
The rendered quantity is the same passive mass fraction tracer shown in Figures \ref{fig:aljet} and \ref{fig:orjet}. 
In this case the jet axis is viewed in the plane of the sky. The
upper image corresponds to a $\mathcal{M}_j = 3.5$ jet pair, while the lower image represents
the cocoons of a $\mathcal{M}_j = 10$ jet pair. The structures
are viewed when $\ell_j \approx 140-150 r_j$. The jets, themselves, are faintly recognizable inside
the cocoons and along the cocoon axes. The cocoon boundaries
are clearly influenced by KHI; some mixing has occurred. Indeed the small differences
between the left and right cocoons come from detailed differences in the KHI
on the two sides that are seeded by the mismatch between a mathematically circular jet and a Cartesian
numerical grid. KHI details depend on the exact placement of the AGN on the numerical grid.
The jet mass fraction dominating the images $\ga 80\%$. Despite such
complications, there is
a relatively clear cocoon boundary, so we assume below that the jet cocoon and the surrounding ICM are cleanly separated.

\begin{figure}
\includegraphics[width=0.48\textwidth,height=0.41\textwidth]{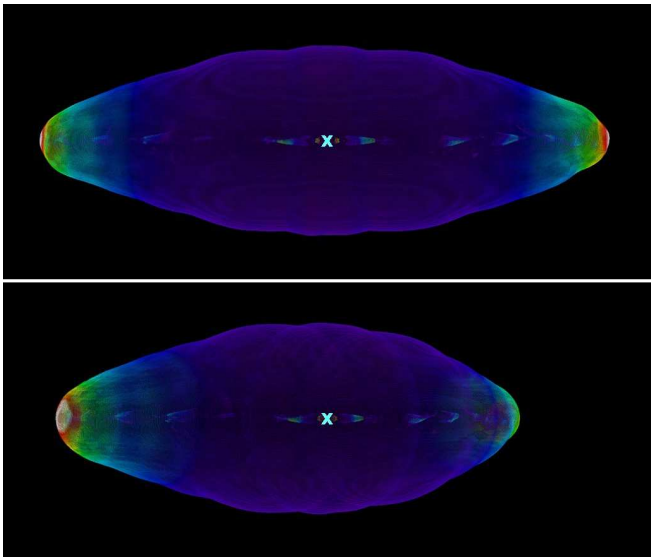}
\caption{\label{fig:noshock-mach} Volume rendering of the shocks associated with the
$\mathcal{M}_j = 10$ simulation shown in the bottom image of 
Figure \ref{fig:noshock-color}. The {\bf{Top}} image matches the orientation of the bottom image in Figure \ref{fig:noshock-color}, while the {\bf{Bottom}}  image has the jet axis rotated 40 degrees, with the left jet approaching (so similar to Figures \ref{fig:aljet} and \ref{fig:orjet}. Shocks are color coded, with the strongest shocks, $\mathcal{M}_s \approx 7$ shown in white, with $\mathcal{M}_s \approx 1.5$ represented by purple.
'X' marks AGN location. See text for details.}
\end{figure}

The energy deposition from the jet into the cocoon will generally drive a shock laterally
outward into the ICM. We will call this the ``inflation shock''.
Provided the jet terminus propagates supersonically into the ICM ($\mathcal{M}_h \mp \mathcal{M}_{w,\parallel} > 1$, equation \ref{eq:headv-mach}) there will be a bow shock attached to the jet head, which will
generally merge with the inflation shock. In Figure \ref{fig:noshock-color} the tips of
the two cocoons are confined by bow shocks. The tighter ``Mach cone'' of the
higher Mach number jet leads to ``sharper points'' in the cocoons. Although the bow shock may be quite strong, both observations and
simulations suggest that the inflation shock is typically rather weak. 
For example the $\mathcal{M}_j = 10$ jets shown in the bottom of
Figure \ref{fig:noshock-color} and in Figure \ref{fig:noshock-mach}
produce bow shocks with $\mathcal{M}_h \approx 7$ ($v_{w,\parallel} = 0$) on the nose
of the jet in agreement with expectations from equation \ref{eq:headv-app}. On the other
hand, over much of their surfaces the accompanying both inflation shocks have Mach
numbers, $\mathcal{M} \la 1.5$ for the duration of the simulation (see Figure \ref{fig:noshock-mach}). There is no
fixed value, however. The relatively small Mach numbers of most of the inflation shock surfaces
also point to the fact that the pressure inside the shocks and inside the cocoons is
not much greater than the ambient pressure, $P_i$ (see equation \ref{eq:jump-p}). 

Because these properties do not inherently lead to scale-free structures, we do not try to model cocoon formation
in terms of self-similar behavior \cite{falle91,ka97}.
Still, it makes sense to estimate the volume of the cocoon simply by comparing the
work required to inflate the cocoon to the energy that has passed down the jet into
the cocoon. Very simply, we set $P_i V_c(t) = K_w \int L_j dt = K_w \langle L_j \rangle t$,
where $K_w < 1$ is a numerical factor that accounts for energy lost by the
cocoon plasma as it inflated. Accounting for adiabatic work done on the ICM would
lead to $K_w = (\gamma_j - 1)/\gamma_j$. For a non-relativistic jet with $\gamma_j = 5/3$, this would give $K_w = 2/5$, while for a jet with a relativistic EoS and $\gamma_j = 4/3$, the equivalent consequences would give $K_w = 1/4$. Empirical
estimates from non-relativistic simulations do, indeed, suggest $K_w \sim 1/2$.\cite{oj10} The exact
value for $K_w$ is not essential to our purposes here.

As a primitive  model for cocoon geometry we assume the cocoon length is set by $\int v_h(t) dt = \langle v_h\rangle t$, giving
\begin{equation}
V_c  = \langle v_h\rangle  A_c \approx \frac{K_w \langle L_j\rangle}{P_i}, 
\label{eq:cvol-simp}
\end{equation}
where $A_c$ represents a characteristic cocoon cross sectional area.
We do not imply in this that the cocoon cross section is constant, and, indeed,
it generally will not be (see Figure \ref{fig:noshock-color}). In practice we treat it as a geometric average cross section.

To simplify our treatment further, we express the jet luminosity in the form applicable
to both relativistic and non-relativistic jets suggested by
equations \ref{eq:ljet-nonrel} and \ref{eq:ljet-rel}; that is,
\begin{equation}
L_j \approx \alpha_j \mathcal{M}_j^2 v_j P_j A_j,
\label{eq:ljet-simp}
\end{equation}
where $1 \le \alpha \le 2$. For a relativistic jet $v_j \to c$ in this expression.
From equation \ref{eq:headv-app} we write
\begin{equation}
v_h \approx  a_i \mathcal{M}_j \sqrt{\frac{A_j}{A_h}}\sqrt{\frac{P_j}{P_i}}\sqrt{\frac{\gamma'}{\gamma_i}}
\label{eq:headv-simp}
\end{equation}
Combining equations \ref{eq:cvol-simp}, \ref{eq:ljet-simp} and \ref{eq:headv-simp} gives 
a simple estimate for $A_c$,
\begin{equation}
A_c \approx \sqrt{A_j A_h}~\mathcal{M}_j \frac{v_j}{a_j} \sqrt{\frac{a_j^2\gamma_i}{a_i^2\gamma'}}\sqrt{\frac{P_j}{P_i}}~(\alpha K_w).
\label{eq:ac-gen}
\end{equation}
We expect $\alpha K_w \la 1$, and, indeed using equation \ref{eq:ac-nonrel}, below, and the simulation parameters for the cocoons shown in
Figure \ref{fig:noshock-color} we get consistent estimates for $A_c$ using
$\alpha K_w \sim 1/2$. Note that, if these various jet parameters are time
independent, the effective cocoon cross section, $A_c$ is steady in time. That is, the ratio $A_c/\ell_j \propto 1/t$.
We briefly address alternative possibilities below.

For a relativistic jet with a relativistic EoS, $v_j/a_j \to \sqrt{3}$, while $\tilde{w}_j \to 4P_j/(\rho_j c^2)$, 
and equation \ref{eq:ac-gen}  becomes,
\begin{eqnarray}
\label{eq:ac-rel}
A_c \approx \sqrt{A_j A_h}~\mathcal{M}_j\left(\sqrt{\frac{\rho_i}{\rho_j}}~\frac{P_j}{P_i}\right)\sqrt{\frac{\rho_j c^2}{2 P_j}}~(\alpha K_w)\\
= \sqrt{A_j A_h}~\mathcal{M}_j\sqrt{\frac{P_j}{P_i}}\sqrt{\gamma_i}\frac{c}{a_i}~(\alpha K_w).\nonumber
\end{eqnarray}
while for a non-relativistic jet,
\begin{equation}
A_c \approx \sqrt{A_j A_h}~\mathcal{M}_j^2\left( \sqrt{\frac{\rho_i}{\rho_j}}~ \frac{P_j}{P_i}\right)~(\alpha K_w).
\label{eq:ac-nonrel}
\end{equation}
The only difference between equations \ref{eq:ac-rel} and \ref{eq:ac-nonrel} is a
relative factor $(1/\mathcal{M}_j)\sqrt{\rho_j c^2/(2 P_j)}$ in equation \ref{eq:ac-rel}.

We can see from these relations that the cocoon cross section scales
with the geometric mean of the jet cross section, $A_j$ and the jet head cross section, $A_h$. It is strongly boosted if the jet Mach number is large (especially for non-relativistic jets), and, with less sensitivity,
to a large density contrast between the ICM and the jet, $\rho_i/\rho_j$. That is,
cocoons will be somewhat fatter for jets with larger density contrast, $\rho_i / \rho_j$, but especially fatter for larger Mach numbers. 
We are assuming in these comments that, as argued previously, we should expect $P_j/P_i \sim 1$.
The Mach number sensitivity of $A_c$ is
evident in a comparison of the non-relativistic jet cocoons shown in Figure \ref{fig:noshock-color}, where the only difference in the two simulations was jet Mach number;  above, $\mathcal{M}_j = 3.5$, while below, $\mathcal{M}_j = 10$. 
The $\mathcal{M}_j = 10$ cocoon, in particular, is, as noted above, strongly tapered, representing
the fact that the Mach cone at its nose has strongly confined it. 
We associate $A_c$ with the mean cross sectional area, which here would be roughly
midway between the AGN and the jet terminus.

The reader may have noticed at the same time that the cocoons from the 
jets shown in Figure \ref{fig:noshock-color} are relatively skinny compared to typical, observed radio galaxy cocoons. 
That is characteristic of the properties of moderate Mach number, non-relativistic AGN jet simulations that are steady and
maintain a precise axis for the jets. On the other hand, if either the jet power cycles substantially 
 or the axis of the jet wanders or precesses, the relative rate at which the
jet advances is reduced \cite{oj10}. In order to maintain the same cocoon volume indicated 
in equation \ref{eq:cvol-simp} the cross section must increase.
This effect can, be roughly accounted for in
equations \ref{eq:headv-simp} - \ref{eq:ac-nonrel} by adjusting the effective
head area, $A_h$. For instance, a precessing jet will balance its thrust
during a precession period over an area, $A_h \sim 4 \pi r_j \ell_j \gg \pi r_j^2$, where $\ell_j = \ell_j(t)$ is the instantaneous
length of the jet. Applied into equation \ref{eq:headv-simp}
the head advance rate, $v_h  = d\ell_j/dt \propto 1/t^{1/2}$. If we maintain constant jet
luminosity, $V_c \propto t$, so $A_c \propto t^{1/2}$. Now, $A_c /\ell_j =$ const, rather
than $A_c/\ell_j \propto 1/t$.
Simultaneously, the volume swept out by the precessing jet itself will be $\sim \ell_j^3 \sin{\theta}\sin{2\theta}$, where $\theta$ is the opening angle of the precession cone.
This sets a scale for the effective $A_c \sim 2~\ell_j^2 \sin{\theta}^2$.
These effects both fatten the cocoons relative to what we derived above.

Finally, we mention that equation \ref{eq:ac-rel} predicts rather fat
cocoons for high Mach number relativistic jets propagating in an environment where $c\gg a_i$.
That is consistent with results of numerical simulations of relativistic
hydrodynamical jets \cite{pm07}, and also
simulations of jets dominated by their Poynting flux (so, internally
highly relativistic) \cite{li06}.

\section{Summary}

The plasma media in galaxy clusters, ICMs, are dynamical. They are stirred by many
processes associated with cluster assembly. The dynamical state of an ICM provides
unique information about how this takes place, so it is important to
find and evaluate ICM dynamical conditions, especially those far away from
equilibrium. X-ray observations can reveal critical information about
ICM thermodynamical properties, some statistical characteristics of
ICM dynamical states and find relatively strong shocks in higher density
ICM regions.

But, much of the story is not visible in the X-ray window. On the other hand, AGNs in so-called ``radio mode'' are common in clusters, and especially in disturbed clusters. Those AGNs  expel fast plasma jets that plow through
the ICM. Those interactions will influence the ICM and its evolution. More to the point of
this paper, however, is the considerable
impact of ICM dynamics on the trajectories and of the AGN jets and the distributions of
their debris. In particular, impacts between ICM shocks and winds will distort
and even disrupt the AGN structures that form. Through an understanding of how
those distortions develop and how they depend on the AGN and ICM properties we
hope to open a clearer window to revealing the dynamical structures of ICMs. By relating
those structures to other information about the dynamical state of the cluster, and,
for instance, evidence for merging, strong accretion events or gravitational
disturbance by dark matter halos, these insights can provide vital probes of
cluster formation processes and their relative roles.

We have laid out in this paper a simple summary of some of the essential dynamical
relationships involved in AGN/ICM interactions, with a special focus on interactions
involving ICM shocks. The nominal target application is ICM shocks colliding with AGN outflows. We pointed out, however, that the relations we developed have application to any relative motion between the AGN and its immediate environment. The formalism allows AGN jets that are
either non-relativistic or relativistic.

We set down basic relations to evaluate
shock interactions with the low density cavities created by AGN jets, as well as to
follow the propagation of the jets within post-shock flows. For completeness, we
used the same formalism to provide a basic context for the formation of the
cavities themselves. One notable aspect of the relationships we derive is the essential
roles of the internal Mach number of the jet flow and the Mach number of the ICM shock.
More directly, the ratio of these two Mach numbers seems central to evaluating the
interactions between the AGN and a dynamical ambient environment.
This provides a potentially useful link that can help develop quantitative
understandings of ICM dynamical states that otherwise are likely to remain
obscure for some time.

\begin{acknowledgments}
TWJ and BO were supported in this work at the University of Minnesota by NSF grant AST1211595. CN was supported by an NSF Graduate Fellowship under Grant 000039202. TWJ, CN and
BO gratefully acknowledge support and hospitality of the Minnesota Supercomputing Insitute. We thank an anonymous referee for constructive comments that improved the manuscript.
\end{acknowledgments}



%
\end{document}